\documentclass[sigconf]{acmart}

\AtBeginDocument{%
  }

\usepackage{hyperref}
\usepackage{enumitem}
\usepackage{makecell}
\usepackage{array}
\usepackage{multirow}

\definecolor{codebgd}{RGB}{255, 253, 232}
\definecolor{lightgray}{RGB}{200, 200, 200}
\definecolor{codehlt}{RGB}{245, 245, 254}
\definecolor{shade1}{RGB}{150,220,0}
\definecolor{shade2}{RGB}{0,0,120}

\newcommand{\ipstart}[1]{\vspace{1mm}\noindent{\textbf{\textit{#1.}}}}

\definecolor{customblue}{rgb}{0.0, 0.5, 1.0}
\definecolor{customlightblue}{rgb}{0.53, 0.81, 0.98}
\definecolor{customgray}{rgb}{0.83, 0.83, 0.83}
\definecolor{custompeach}{rgb}{0.98, 0.8, 0.69}
\definecolor{customorange}{rgb}{0.95, 0.55, 0.2}
\definecolor{custommidorangepeach}{rgb}{0.965, 0.675, 0.445}
\definecolor{cmumaroon}{rgb}{192, 0, 0}
\definecolor{shade1}{RGB}{192,0,0}
\definecolor{shade2}{RGB}{18,96,157}
\definecolor{custommidbluelightblue}{rgb}{0.265, 0.655, 0.99}
\definecolor{neongreen}{rgb}{0.1, 1.0, 0.1}
\definecolor{neonyellow}{rgb}{0.98, 1.0, 0.18}
\newenvironment{highlight}
  {\color{black}}
  {}            
\newcommand{\likertpct}[8][0.225]{%
\begin{tabular}{rcl}
\the\numexpr(#4+#2+#3)*100/16\relax\% &
\resizebox{#1\textwidth}{0.78\height}{%
\color{customorange}\rule{#2mm}
{10pt}\color{custommidorangepeach}\rule{#3mm}{10pt}\color{custompeach}\rule{#4mm}{10pt}\color{customgray}\rule{#5mm}{10pt}\color{customlightblue}\rule{#6mm}
{10pt}\color{custommidbluelightblue}\rule{#7mm}{10pt}\color{customblue}\rule{#8mm}{10pt}%
} &
\the\numexpr(#6+#7+#8)*99/16\relax\%\\
\end{tabular}%
}

\newcommand{\likertpctt}[8][0.225]{%
\begin{tabular}{rcl}
\the\numexpr(#4+#2+#3)*5\relax\% &
\resizebox{#1\textwidth}{0.78\height}{%
\color{customorange}\rule{#2mm}
{10pt}\color{custommidorangepeach}\rule{#3mm}{10pt}\color{custompeach}\rule{#4mm}{10pt}\color{customgray}\rule{#5mm}{10pt}\color{customlightblue}\rule{#6mm}
{10pt}\color{custommidbluelightblue}\rule{#7mm}{10pt}\color{customblue}\rule{#8mm}{10pt}%
} &
\the\numexpr(#6+#7+#8)*5\relax\%\\
\end{tabular}%
}

\newcommand{\likerteval}[8][0.325]{%
\begin{tabular}{rcl}
\the\numexpr(#4+#2+#3)*5\relax\% &
\resizebox{#1\textwidth}{0.78\height}{%
\color{customorange}\rule{#2mm}{10pt}\color{custommidorangepeach}\rule{#3mm}{10pt}\color{custompeach}\rule{#4mm}{10pt}\color{customgray}\rule{#5mm}{10pt}\color{customlightblue}\rule{#6mm}
{10pt}\color{custommidbluelightblue}\rule{#7mm}{10pt}\color{customblue}\rule{#8mm}{10pt}%
} &
\the\numexpr(#6+#7+#8)*5\relax\%\\
\end{tabular}%
}
\copyrightyear{2025}
\acmYear{2025}
\setcopyright{rightsretained}
\acmConference[CHI '25]{CHI Conference on Human Factors in Computing Systems}{April 26-May 1, 2025}{Yokohama, Japan}
\acmBooktitle{CHI Conference on Human Factors in Computing Systems (CHI '25), April 26-May 1, 2025, Yokohama, Japan}\acmDOI{10.1145/3706598.3713335}
\acmISBN{979-8-4007-1394-1/25/04}

\begin{document}

\title{CodeA11y: Making AI Coding Assistants Useful for Accessible Web Development}






\author{Peya Mowar}
\email{pmowar@cs.cmu.edu}
\affiliation{%
  \institution{Carnegie Mellon University}
  \city{Pittsburgh}
  \state{PA}
  \country{USA}}

\author{Yi-Hao Peng}
\email{yihaop@cs.cmu.edu}
\affiliation{%
  \institution{Carnegie Mellon University}
  \city{Pittsburgh}
  \state{PA}
  \country{USA}}

\author{Jason Wu}
\email{jason_wu8@apple.com}
\affiliation{%
  \institution{Apple}
  \city{Seattle}
  \state{WA}
  \country{USA}}

\author{Aaron Steinfeld}
\email{steinfeld@cmu.edu}
\affiliation{%
  \institution{Carnegie Mellon University}
  \city{Pittsburgh}
  \state{PA}
  \country{USA}}

\author{Jeffrey P. Bigham}
\email{jbigham@cs.cmu.edu}
\affiliation{%
  \institution{Carnegie Mellon University}
  \city{Pittsburgh}
  \state{PA}
  \country{USA}}









\label{abstract}
\begin{abstract}
A persistent challenge in accessible computing is ensuring developers produce web UI code that supports assistive technologies. Despite numerous specialized accessibility tools, novice developers often remain unaware of them, leading to \textasciitilde 96\% of web pages that contain accessibility violations. 
AI coding assistants, such as GitHub Copilot, could offer potential by generating accessibility-compliant code, but their impact remains uncertain~\cite{mowar2024tab}. Our formative study with 16 developers without accessibility training revealed three key issues in AI-assisted coding: failure to prompt AI for accessibility, omitting crucial manual steps like replacing placeholder attributes, and the inability to verify compliance.
To address these issues, we developed CodeA11y, a GitHub Copilot Extension, that suggests accessibility-compliant code and displays manual validation reminders. We evaluated it through a controlled study with another 20 novice developers. Our findings demonstrate its effectiveness in guiding novice developers by reinforcing accessibility practices throughout interactions, representing a significant step towards integrating accessibility into AI coding assistants.
\end{abstract}





\begin{teaserfigure}
    \includegraphics[width=\textwidth]{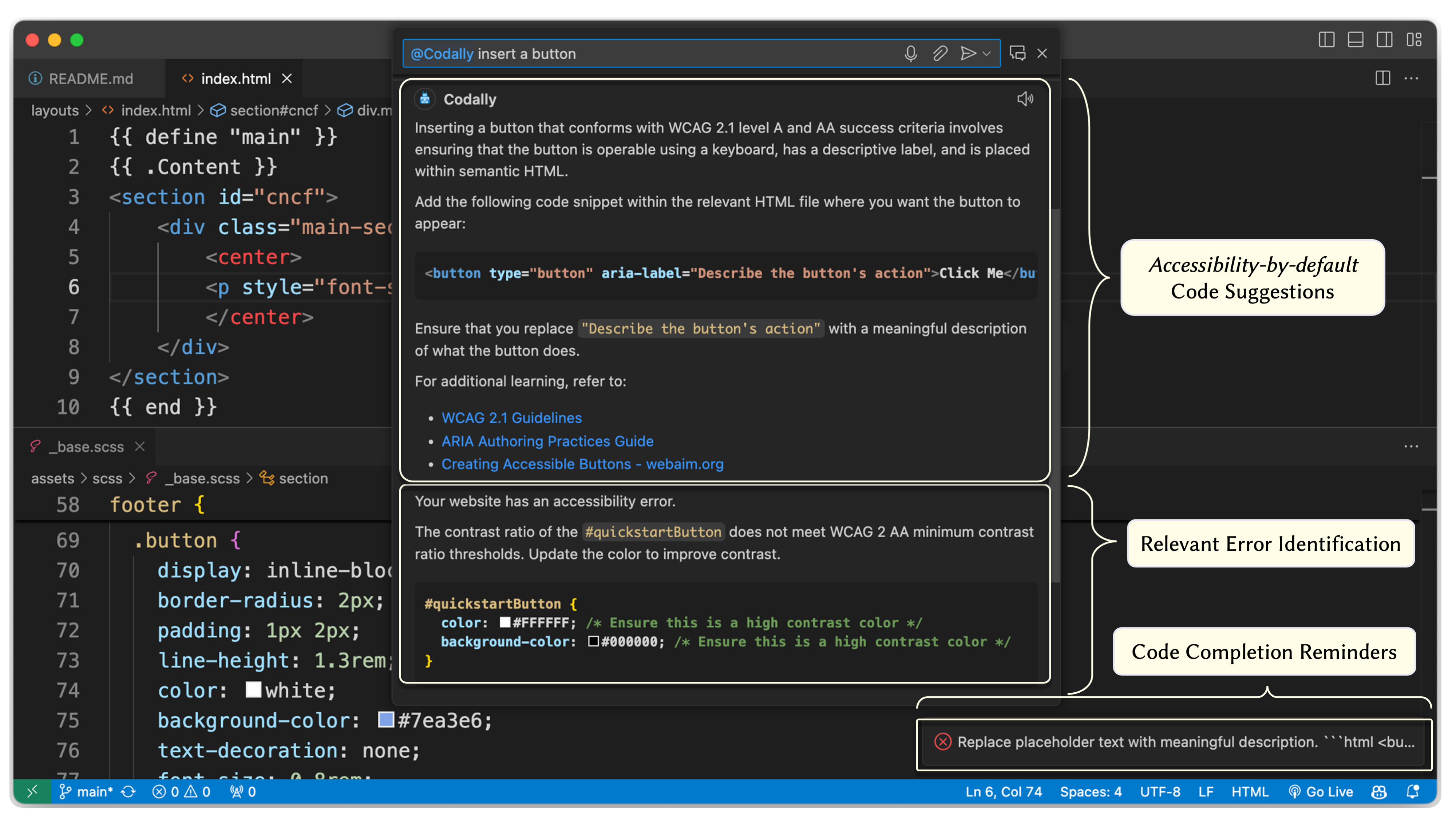}
    \caption{CodeA11y is a GitHub Copilot Extension for Accessible Web Development. CodeA11y addresses accessibility limitations of Copilot observed in our study with developers through three features: (1) accessibility-by-default code suggestions, (2) automatic identification of relevant accessibility errors, and (3) reminders to replace placeholders in generated code. Integrating these features directly into AI coding assistants would improve the accessibility of the user interfaces (UIs) developers create.
    }
    \Description{The interface of the extension is displayed along with its features: (1) Accessibility-by-default Code Suggestions, (2) Relevant Error Identification and (3) Code Completion Reminders.}
    \label{fig:teaser}
\end{teaserfigure}

\maketitle

\title{CodeA11y: Making AI Coding Assistants Useful for Accessible Web Development}

\label{intro}
\section{Introduction}

Most websites contain extensive accessibility errors~\cite{webaim2024}, despite decades of investment in standards and guidelines~\cite{chisholm2001web, caldwell2008web}, tools~\cite{adesigner, takagi2003accessibility, bigham2010accessibility}, advocacy \cite{sloan2006contextual, martin2022landscape, pandey2023blending}, and policy. According to a recent analysis by WebAim~\cite{webaim2024}, the homepages of the top million websites each contain 57 accessibility errors on average, including (but not limited to) missing alt-text for images~\cite{webinsight,twitterally}, inadequate color contrast~\cite{colors}, incorrect or missing labels for forms and links~\cite{formlabels}, and improper use of heading levels~\cite{headings}. As a result, many people with disabilities will find it difficult to use these websites effectively and may not be able to use them at all.

Front-end web developers ultimately determine the accessibility (or inaccessibility) of the UI code that they write. Getting front-end developers to write more accessible code has proven exceptionally difficult. As Jonathan Lazar {\em et al.} wrote twenty years ago in 2004, ``Since tools and guidelines are available to help designers and webmasters make their web sites accessible, it is unclear why so many sites remain inaccessible.''~\cite{lazar2004improving}. A survey of webmasters at the time indicated that they generally would like to make their web pages accessible but cited a number of reasons they do not: "lack of time, lack of training, lack of managerial support, lack of client
support, inadequate software tools, and confusing accessibility guidelines." Sixteen years later, Patel {\em et al.} reported remarkably similar results in their 2020 survey of 77 technology professionals \cite{stillinaccessible}. Few developers had received formal accessibility training, implementing accessibility was considered confusing, and advocating for accessible development was in conflict with other business goals. Clearly, what we have done so far is not working.

We argue that AI coding assistants ({\em e.g.}, Github Copilot~\cite{GitHubCopilot}) could offer an opportunity to make UI code more accessible. AI coding assistants are already widely adopted, which means that developers do not need to be convinced to use them or to install a specialized tool for accessibility. They produce a wide variety of UI code and are capable enough to both reflect on the quality of arbitrary code and also prompt developers to fix what they are unable to do. This paper explores how AI coding assistants currently help developers create UI code, what problems remain, and presents a system called CodeA11y that shows that AI coding assistants can be made better at enabling developers to improve the accessibility of their UI code.

To explore this potential opportunity, we first conducted a user study (Section \ref{form_methods}) with 16 developers not trained in accessibility to understand how current tools (GitHub Copilot) impact the production of accessible UI code. Our findings (Section \ref{form_finds}) shows that while Copilot may potentially improve accessibility of UI code, three barriers prevent realization of those improvements: (1) developers may need to explicitly prompt the assistants for accessible code and thus not benefit if they fail to do so, (2) developers may overlook critical manual steps suggested by Copilot, such as replacing placeholders in alternative text for images, and (3) developers may not be able to verify if they fully implemented more complex accessibility enhancements properly. The formative study showed the potential of AI coding assistants to improve the accessibility of UI code, but revealed several gaps that led us to design goals (Section~\ref{design-goals}) for improving AI coding assistants to support accessibility.

We then built \textit{CodeA11y} (Section~\ref{system}, Figure~\ref{fig:teaser}), a GitHub Copilot Extension that addresses the observed gaps by consistently reinforcing accessible development practices throughout the conversational interaction. We evaluated CodeA11y (Section~\ref{eval}) with 20 developers, assessing its effectiveness in supporting accessible UI development and gathering insights for further refinement. We found that developers using CodeA11y are significantly more likely to produce accessible UI code. Finally, we reflect on the broader implications of integrating AI coding assistants into accessibility workflows, including the balance between automation and developer education, and the potential for AI tools to shape long-term developer behavior toward accessibility-conscious practices (Section~\ref{discuss}).

The contributions of our paper are:
\begin{itemize}[noitemsep, topsep=0pt]
    \item We conducted a study with 16 developers that uncovered both benefits and limitations of current AI coding assistants for authoring accessible UI code.  
    \item \textbf{CodeA11y\footnote{The source code for CodeA11y is available at \url{https://github.com/peyajm29/codea11y/}.}}: a GitHub Copilot Extension that generates accessible UI code, identifies existing issues and reminds developers to perform manual validation.
\end{itemize}

\section{Related Work}
\label{lit_review}

\begin{highlight}
{

Our research builds upon {\em (i)} Assessing Web Accessibility, {\em (ii)} End-User Accessibility Repair, and {\em (iii)} Developer Tools for Accessibility.

\subsection{Assessing Web Accessibility}
From the earliest attempts to set standards and guidelines, web accessibility has been shaped by a complex interplay of technical challenges, legal imperatives, and educational campaigns. Over the past 25 years, stakeholders have sought to improve digital inclusion by establishing foundational standards~\cite{chisholm2001web, caldwell2008web}, enforcing legal obligations~\cite{sierkowski2002achieving, yesilada2012understanding}, and promoting a broader culture of accessibility awareness among developers~\cite{sloan2006contextual, martin2022landscape, pandey2023blending}. 
Despite these longstanding efforts, systemic accessibility issues persist. According to the 2024 WebAIM Million report~\cite{webaim2024}, 95.9\% of the top one million home pages contained detectable WCAG violations, averaging nearly 57 errors per page. 
These errors take many forms: low color contrast makes the interface difficult for individuals with color deficiency or low vision to read text; missing alternative text leaves users relying on screen readers without crucial visual context; and unlabeled form inputs or empty links and buttons hinder people who navigate with assistive technologies from completing basic tasks. 
Together, these accessibility issues not only limit user access to critical online resources such as healthcare, education, and employment but also result in significant legal risks and lost opportunities for businesses to engage diverse audiences. Addressing these pervasive issues requires systematic methods to identify, measure, and prioritize accessibility barriers, which is the first step toward achieving meaningful improvements.

Prior research has introduced methods blending automation and human evaluation to assess web accessibility. Hybrid approaches like SAMBA combine automated tools with expert reviews to measure the severity and impact of barriers, enhancing evaluation reliability~\cite{brajnik2007samba}. Quantitative metrics, such as Failure Rate and Unified Web Evaluation Methodology, support large-scale monitoring and comparative analysis, enabling cost-effective insights~\cite{vigo2007quantitative, martins2024large}. However, automated tools alone often detect less than half of WCAG violations and generate false positives, emphasizing the need for human interpretation~\cite{freire2008evaluation, vigo2013benchmarking}. Recent progress with large pretrained models like Large Language Models (LLMs)~\cite{dubey2024llama,bai2023qwen} and Large Multimodal Models (LMMs)~\cite{liu2024visual, bai2023qwenvl} offers a promising step forward, automating complex checks like non-text content evaluation and link purposes, achieving higher detection rates than traditional tools~\cite{lopez2024turning, delnevo2024interaction}. Yet, these large models face challenges, including dependence on training data, limited contextual judgment, and the inability to simulate real user experiences. These limitations underscore the necessity of combining models with human oversight for reliable, user-centered evaluations~\cite{brajnik2007samba, vigo2013benchmarking, delnevo2024interaction}. 

Our work builds on these prior efforts and recent advancements by leveraging the capabilities of large pretrained models while addressing their limitations through a developer-centric approach. CodeA11y integrates LLM-powered accessibility assessments, tailored accessibility-aware system prompts, and a dedicated accessibility checker directly into GitHub Copilot---one of the most widely used coding assistants. Unlike standalone evaluation tools, CodeA11y actively supports developers throughout the coding process by reinforcing accessibility best practices, prompting critical manual validations, and embedding accessibility considerations into existing workflows.

\subsection{End-user Accessibility Repair}
In addition to detecting accessibility errors and measuring web accessibility, significant research has focused on fixing these problems.
Since end-users are often the first to notice accessibility problems and have a strong incentive to address them, systems have been developed to help them report or fix these problems.

Collaborative, or social accessibility~\cite{takagi2009collaborative,sato2010social}, enabled these end-user contributions to be scaled through crowd-sourcing.
AccessMonkey~\cite{bigham2007accessmonkey} and Accessibility Commons~\cite{kawanaka2008accessibility} were two examples of repositories that store accessibility-related scripts and metadata, respectively.
Other work has developed browser extensions that leverage crowd-sourced databases to automatically correct reading order, alt-text, color contrast, and interaction-related issues~\cite{sato2009s,huang2015can}.

One drawback of collaborative accessibility approaches is that they cannot fix problems for an ``unseen'' web page on-demand, so many projects aim to automatically detect and improve interfaces without the need for an external source of fixes.
A large body of research has focused on making specific web media (e.g., images~\cite{gleason2019making,guinness2018caption, twitterally, gleason2020making, lee2021image}, design~\cite{potluri2019ai,li2019editing, peng2022diffscriber, peng2023slide}, and videos~\cite{pavel2020rescribe,peng2021say,peng2021slidecho,huh2023avscript}) accessible through a combination of machine learning (ML) and user-provided fixes.
Other work has focused on applying more general fixes across all websites.

Opportunity accessibility addressed a common accessibility problem of most websites: by default, content is often hard to see for people with visual impairments, and many users, especially older adults, do not know how to adjust or enable content zooming~\cite{bigham2014making}.
To this end, a browser script (\texttt{oppaccess.js}) was developed that automatically adjusted the browser's content zoom to maximally enlarge content without introducing adverse side-effects (\textit{e.g.,} content overlap).
While \texttt{oppaccess.js} primarily targeted zoom-related accessibility, recent work aimed to enable larger types of changes, by using LLMs to modify the source code of web pages based on user questions or directives~\cite{li2023using}.

Several efforts have been focused on improving access to desktop and mobile applications, which present additional challenges due to the unavailability of app source code (\textit{e.g.,} HTML).
Prefab is an approach that allows graphical UIs to be modified at runtime by detecting existing UI widgets, then replacing them~\cite{dixon2010prefab}.
Interaction Proxies used these runtime modification strategies to ``repair'' Android apps by replacing inaccessible widgets with improved alternatives~\cite{zhang2017interaction, zhang2018robust}.
The widget detection strategies used by these systems previously relied on a combination of heuristics and system metadata (\textit{e.g.,} the view hierarchy), which are incomplete or missing in the accessible apps.
To this end, ML has been employed to better localize~\cite{chen2020object} and repair UI elements~\cite{chen2020unblind,zhang2021screen,wu2023webui,peng2025dreamstruct}.

In general, end-user solutions to repairing application accessibility are limited due to the lack of underlying code and knowledge of the semantics of the intended content.

\subsection{Developer Tools for Accessibility}
Ultimately, the best solution for ensuring an accessible experience lies with front-end developers. Many efforts have focused on building adequate tooling and support to help developers with ensuring that their UI code complies with accessibility standards.

Numerous automated accessibility testing tools have been created to help developers identify accessibility issues in their code: i) static analysis tools, such as IBM Equal Access Accessibility Checker~\cite{ibm2024toolkit} or Microsoft Accessibility Insights~\cite{accessibilityinsights2024}, scan the UI code's compliance with predefined rules derived from accessibility guidelines; and ii) dynamic or runtime accessibility scanners, such as Chrome Devtools~\cite{chromedevtools2024} or axe-Core Accessibility Engine~\cite{deque2024axe}, perform real-time testing on user interfaces to detect interaction issues not identifiable from the code structure. While these tools greatly reduce the manual effort required for accessibility testing, they are often criticized for their limited coverage. Thus, experts often recommend manually testing with assistive technologies to uncover more complex interaction issues. Prior studies have created accessibility crawlers that either assist in developer testing~\cite{swearngin2024towards,taeb2024axnav} or simulate how assistive technologies interact with UIs~\cite{10.1145/3411764.3445455, 10.1145/3551349.3556905, 10.1145/3544548.3580679}.

Similar to end-user accessibility repair, research has focused on generating fixes to remediate accessibility issues in the UI source code. Initial attempts developed heuristic-based algorithms for fixing specific issues, for instance, by replacing text or background color attributes~\cite{10.1145/3611643.3616329}. More recent work has suggested that the code-understanding capabilities of LLMs allow them to suggest more targeted fixes.
For example, a study demonstrated that prompting ChatGPT to fix identified WCAG compliance issues in source code could automatically resolve a significant number of them~\cite{othman2023fostering}. Researchers have sought to leverage this capability by employing a multi-agent LLM architecture to automatically identify and localize issues in source code and suggest potential code fixes~\cite{mehralian2024automated}.

While the approaches mentioned above focus on assessing UI accessibility of already-authored code (\textit{i.e.,} fixing existing code), there is potential for more proactive approaches.
For example, LLMs are often used by developers to generate UI source code from natural language descriptions or tab completions~\cite{chen2021evaluating,GitHubCopilot,lozhkov2024starcoder,hui2024qwen2,roziere2023code,zheng2023codegeex}, but LLMs frequently produce inaccessible code by default~\cite{10.1145/3677846.3677854,mowar2024tab}, leading to inaccessible output when used by developers without sufficient awareness of accessibility knowledge.
The primary focus of this paper is to design a more accessibility-aware coding assistant that both produces more accessible code without manual intervention (\textit{e.g.,} specific user prompting) and gradually enables developers to implement and improve accessibility of automatically-generated code through IDE UI modifications (\textit{e.g.}, reminder notifications).

}
\end{highlight}

\section{Formative Study Methods}
\label{form_methods}
We conducted a formative study to assess the implications of AI coding assistants on web accessibility. We recruited novice developers and tasked them with editing real-world websites using GitHub Copilot. Our goal was to better understand how the use of Copilot affected the accessibility of the user interface code they produced.


\ipstart{Tasks}
The participants completed tasks in the codebases for two open-source websites, Kubernetes~\cite{kubernetes} and BBC News~\cite{bbcnews}. Both websites received over 2 million monthly visits worldwide~\cite{similarweb2024} and belong to different categories in the IAB Content Taxonomy~\cite{webshrinker2024}. These websites were developed using different web development frameworks (Hugo and React, respectively). To choose the four specific tasks used in this formative study (Table~\ref{tab:tasks}), we sampled actual issues from each website's repository. We chose issues for which accessibility needed to be considered to complete them correctly, but accessibility was not explicitly mentioned as a requirement in either the task description given to participants or on the issue description on the website's code repository, as illustrated in Figure~\ref{fig:tasks}. Correctly performing the tasks required the consideration of several common web accessibility issues: color contrast, alternative text, link labels, and form labeling~\cite{webaim2024}. The goal was to mirror the kinds of specifications that developers often receive that do not explicitly mention accessibility.


\begin{highlight}
\begin{figure*}
    \includegraphics[width=\textwidth, trim=15 280 12 100, clip]{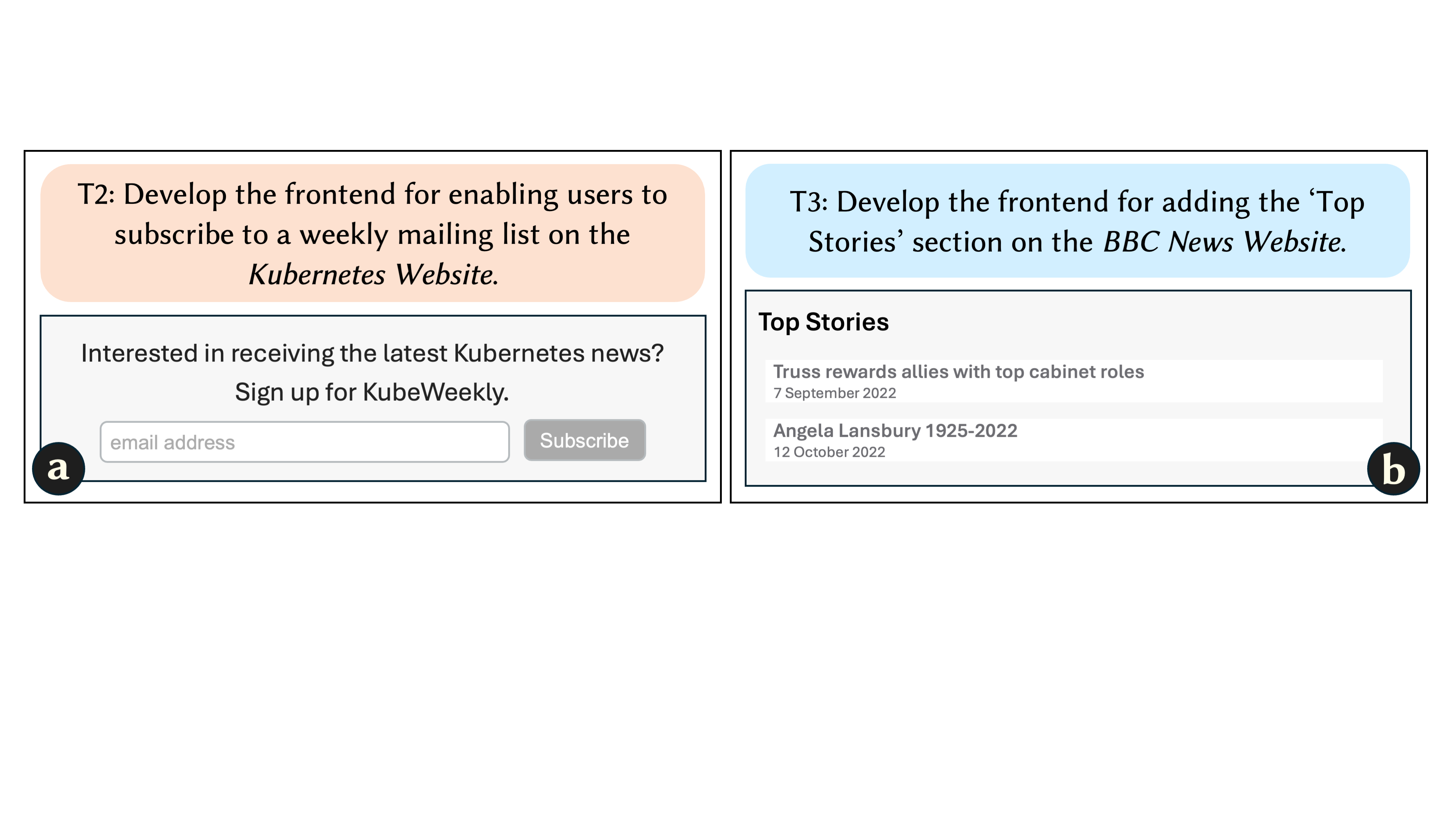}
    \caption{\begin{highlight}Examples of task descriptions and visual references given to our participants: (a) Task 2 was to implement a new contact form for subscribing to a mailing list, and (b) Task 3 was to add a `Top Stories' section with linked articles. Successfully completing them required proper labeling of the form elements and links, but this was not explicitly stated in the instructions.\end{highlight}}
    \Description{The figure depicts the two tasks: i) Task T2's instruction is about developing a subscription feature for a weekly mailing list on the Kubernetes Website. It's visual reference includes a subscription box labeled "email address" with a "Subscribe" button, inviting users to sign up for "KubeWeekly" updates. ii) Task T3's instruction is to develop the frontend for adding the Top Stories section on the BBC Website. t's visual reference includes a list of headlines with corresponding dates styled as clickable elements.}
    \label{fig:tasks}
\end{figure*}
\end{highlight}

\begin{table*}
  \caption{Our formative study included four tasks. Each task was not primarily about accessibility but included an accessibility issue that was required to complete the task successfully. \begin{highlight} We adopt the scales of Unacceptable, Average and Good from prior work~\cite{pillai2022website}. Uninformative attributes are those that merely reflect the field, such as `alt' as alt-text or `click here' as link description, without providing more meaningful or descriptive content~\cite{ross2018examining}. Tasks are ranked from easy to difficult based on the time taken and success rates observed in our pilot studies.\end{highlight}}
  \label{tab:tasks}
  \small
  \begin{tabular}{|p{0.185\textwidth}|p{0.075\textwidth}|p{0.13\textwidth}|p{0.525\textwidth}|}
    \toprule
    \textbf{Task} & \textbf{Difficulty} & \textbf{Accessibility Issue} & \textbf{Evaluation Criteria}\\
    \midrule
    (T1) Button Visibility & Easy & Color Contrast & \textit{Unacceptable}: contrast ratio of $<$ 4.5:1 for normal text and $<$ 3:1 for large text\\
    & & & \textit{Average}: WCAG level AA in default state (contrast ratio of $>=$ 4.5:1 for normal text) \\
    & & & \textit{Good}: WCAG level AA in all states (default, hover, active, focus, etc.)\\
    \midrule
    (T2) Form Element & Moderate & Form Labeling & \textit{Unacceptable}: Missing form labels and keyboard navigation \\
    & & & \textit{Average}: One of form labels and keyboard navigation\\
    & & & \textit{Good}: Both form labeling and keyboard navigation \\
    \midrule
    (T3) Add Section & Moderate & Link Labeling & \textit{Unacceptable}: Missing link descriptions \\
    & & & \textit{Average}: Uninformative link descriptions \\
    & & & \textit{Good}: Descriptive link descriptions\\
    \midrule
    (T4) Enhance Image for SEO & Difficult & Adding alt-text & \textit{Unacceptable}: Missing or uninformative alt-text\\
     & & & \textit{Average}: Added alt-text with $<$ 3 required descriptors~\cite{10.1145/3441852.3471207}\\
    & & & \textit{Good}: Added alt-text with $>=$ 3 out of 4 required descriptors\\
    \bottomrule
  \end{tabular}
\end{table*}

\begin{table}[h!]
\caption{\begin{highlight}
Participant User Groups: Each group is assigned specific order of tasks and testing conditions. Participants are evenly and randomly distributed among these groups.
\end{highlight}}
\centering
\small
\label{tab:usergroup}
\begin{highlight}
\begin{tabular}{|p{0.01\textwidth}|p{0.2\textwidth}|p{0.2\textwidth}|}
\toprule
\textbf{\#} & \textbf{Order 1, Testing Condition} & \textbf{Order 2, Testing Condition} \\ \midrule
1 & Kubernetes, With AI Assistance & BBC News, No AI Assistance \\ \midrule
2 & Kubernetes, No AI Assistance & BBC News, With AI Assistance \\ \midrule
3 & BBC News, With AI Assistance & Kubernetes, No AI Assistance \\ \midrule
4 & BBC News, No AI Assistance & Kubernetes, With AI Assistance \\ \midrule
\end{tabular}
\end{highlight}
\end{table}

\ipstart{Protocol}
Our within-subjects user study had two conditions: (1) a control condition where participants received no AI assistance, and (2) a test condition where participants used GitHub Copilot. Each participant was assigned to edit two distinct websites, each with two tasks. To counterbalance order effects, participants were evenly and randomly assigned to one of four user groups (Table~\ref{tab:usergroup}), balanced by website order and control/test conditions. Further, to simulate real-world scenarios, we concealed the true purpose of the study from participants. Participants were informed that the study was about the usability of AI pair programmers in web development tasks but were not explicitly instructed to make their web components accessible. This allowed us to observe how developers naturally handle accessibility when it is not explicitly emphasized, reflecting typical developer behavior. The research protocol was reviewed and approved by the Institutional Review Board (IRB) at our university.

\ipstart{Participants}
We employed convenience sampling and snowball sampling methods to recruit our participants. Our study was advertised on university bulletin boards, social media, and shared communication channels (Twitter, Slack, and mailing groups). Our recruitment criteria stipulated that participants must be over 18 years of age, live in the United States, and have self-assessed familiarity with web development. Further, we required the participants to be physically present on our university campus for the duration of the study. To avoid priming during participant recruitment, we did not stipulate awareness of web accessibility as an eligibility criterion. We chose university-specific avenues for recruiting CS students, that reflect a typical novice developer cohort.

Our study enlisted 16 participants (7 female and 9 male; ages ranged from 22 to 29). Almost all of our participants were students and had multiple years of coding experience. Most (n=10) had multi-year \textit{industrial} programming experience (e.g., full-time or intern experience in the company). Nearly all participants (except one) had previously used AI coding assistants. GitHub Copilot and OpenAI ChatGPT were the most popular (n=10). Others preferred Tabnine (n=6) and AWS CodeWhisperer (n=2). 12 participants had self-described substantial experience with HTML and CSS. 10 were proficient in JavaScript and 7 were proficient in React.js. Despite this expertise, the majority (14 participants) were unfamiliar with the Web Content Accessibility Guidelines (WCAG). Only 2 participants knew about these guidelines, but they had not actively engaged in creating accessible web user interfaces or received formal training on the subject (details are provided in Table~\ref{tab:awareness}).

\begin{table*}
  \caption{The distribution of participants' opinions on AI-powered programming tools and their awareness of web accessibility. The percentages in the distribution column indicate the proportion of participants who either disagree (including both `strongly disagree' and `disagree') or agree (including both `strongly agree' and `agree') with the provided statements.}
  \label{tab:awareness}
  \begin{tabular}{>{\raggedright\arraybackslash}p{0.6\textwidth}|>{\raggedright\arraybackslash}p{0.35\textwidth}}
    \toprule
    \textbf{Statement} & \textbf{Distribution}\\
    \midrule
    ``I trust the accuracy of AI programming tools.''& \likertpct{0}{0}{2}{10}{4}{0}{0}\\
    ``I am proficient in web accessibility.''& \likertpct{9}{0}{3}{1}{2}{0}{1}\\
    ``I am familiar with the web accessibility standards, such as WCAG 2.0.''& \likertpct{11}{0}{3}{0}{0}{0}{2}\\
    ``I am familiar with ARIA roles, states, and properties.''& \likertpct{10}{0}{1}{1}{3}{0}{1}\\
  \midrule
\end{tabular}
\begin{tabular}{@{}>{\centering\arraybackslash}p{\textwidth}@{}}
        \textcolor{customorange}{\rule{7pt}{7pt}} Strongly Disagree \quad
        \textcolor{custompeach}{\rule{7pt}{7pt}} Disagree \quad
        \textcolor{customgray}{\rule{7pt}{7pt}} Neutral \quad
        \textcolor{customlightblue}{\rule{7pt}{7pt}} Agree \quad
        \textcolor{customblue}{\rule{7pt}{7pt}} Strongly Agree \\
    \bottomrule
    \end{tabular}
\end{table*}

\ipstart{Procedure}
The study was conducted in person at our lab, where participants performed programming tasks on a MacBook Pro laptop equipped with IntelliJ IDEA with the GitHub Copilot plugin preinstalled. Before starting the study, we explained the study procedure to the participants and took their informed consent. The participants then watched a 5-minute instructional video explaining Copilot's features, such as code autocompletion and the Copilot chat\footnote{\url{https://www.youtube.com/watch?v=jXp5D5ZnxGM}}. Participants were assigned tasks related to two selected websites, with a total of four tasks to complete in 90 minutes. They were required to work on one website with and the other without GitHub Copilot. Further, they were allowed to access the web for task exploration or code documentation through traditional search engines like Google Search, but with generative results turned off. During the coding session, a researcher observed silently, offering help with tasks, tool usage, or debugging only if participants were stuck, and asked them to move on after 30 minutes, without giving any accessibility-related hints. Based on our observations from pilot studies, we set time limits ranging from 15 to 30 minutes per task. Finally, after completing the coding tasks, they were asked to participate in a 10-15 minute survey on their experience in AI-assisted programming and web accessibility, development expertise, and open-ended feedback. In the end, the participants were compensated with a gift voucher worth 30 USD.

\ipstart{Data Collection and Analysis}
We collected both quantitative and qualitative data for a mixed-method analysis.
For quantitative data, we used an IntelliJ IDEA plugin~\cite{dkandalov_activity_tracker} that tracked user actions --- such as keyboard input (typing, backspace), IDE commands (copy, paste, undo), and interactions with GitHub Copilot (accepting suggestions, opening the Copilot Chat window) --- and recorded their timestamps. 
Additionally, we employed the axe-Core Accessibility Engine 2 to gather accessibility violation metrics, including the type and count of WCAG failures, for each code submission, a method proven reliable in previous studies~\cite{p2023towards}. We also collected AI usage, programming languages and framework preferences, and expertise in web accessibility via a post-task survey.

On the qualitative side, we captured the entire study sessions through screen recordings, resulting in a total of 18.73 hours of video data. We complemented this with observational notes taken during the sessions, documenting verbal comments made by participants. The participants' interactions with Copilot Chat were also recorded for further analysis between prompts and the final code.
The analysis of this data was carried out using open coding and thematic analysis~\cite{clarke2017thematic}. \begin{highlight} Some themes that emerged were: `visual enhancement', `recalling syntax', `feature request', and `code understanding'.\end{highlight} For accessibility evaluation, we manually inspected the websites created during the study and evaluated their accessibility on a qualitative scale of ``Unacceptable,'' ``Average,'' and ``Good'' adopted from prior work~\cite{pillai2022website}. The criteria for these evaluations were developed per best practices identified in prior research published in CHI and ASSETS, detailed further in Table~\ref{tab:tasks}.

\section{Formative Findings}
\label{form_finds}
Our formative study revealed that while existing AI coding assistants can produce accessible code, developers still need accessibility expertise for effective use. Otherwise, 
(1) the accessibility introduced is likely to not be applied comprehensively, (2) the advanced features recommended by the assistant are unlikely to be implemented,  (3) the accessibility errors introduced by the assistant are unlikely to be caught.

\ipstart{Developer Behavior}
In the study, participants spent slightly more time on tasks without Copilot, averaging $30.84$ minutes ($\sigma = 11.95$) compared to $28.94$ minutes ($\sigma = 8.57$) with Copilot. Copilot also facilitated a greater volume of code edits ($13.28$ lines of code, $\sigma = 9.02$ vs $10.41$ lines of code, $\sigma = 5.87$), indicating that AI-assisted workflows encouraged iterative coding practices. However, even with Copilot, participants spent approximately 39.84\% of their task time ($11.91$ minutes, $\sigma = 8.00$) away from the IDE, browsing the web or checking the rendered HTML, highlighting the importance of traditional validation methods. The study also found fewer backspace key presses, an indicator of post-paste corrections, without Copilot ($\mu = 92.62$, $\sigma = 68.27$) than with the AI assistant ($\mu = 104.50$, $\sigma = 91.91$). Further, code pasting was slightly higher when participants solely browsed the web for exploration, averaging $12.68$ times ($\sigma = 8.09$), compared to $11.43$ times ($\sigma = 5.35$) with access to Copilot Chat. Participants dedicated about $7.39$\% ($2.14$ minutes, $\sigma = 1.72$) of their task time typing in the GitHub Copilot chat window, while, they also accepted Copilot's code auto-complete suggestions around $5.44$ times ($\sigma = 5.00$) on average.

\begin{table*}
\caption{The (partial) chat history revealed that directly copying Copilot's code suggestions would be incomplete, as developers overlooked the additional recommended steps.}
  \small
  \begin{tabular}{p{0.06\textwidth}p{0.89\textwidth}}
    \toprule
    \textbf{User}: & \textcolor{customorange}{``\textbf{User Query}''} \\
    \textbf{Copilot}: &  \colorbox{codebgd}{\texttt{\textbf{Generated Code}}}\\ 
    & \colorbox{codehlt}{\textbf{Associated Text Explanation}}\\
    \midrule
    \textbf{User 1}: & \textcolor{customorange}{\textbf{``I want to label the images or add relevant image descriptors''}} \\
    \textbf{Copilot:} & \colorbox{codebgd}{\texttt{alt={imgAlt} // Add this line}} \\
    & \colorbox{codehlt}{\begin{minipage}{\linewidth}
    Remember to replace imageAlt, <...> with your actual image attributes. In this example, imageAlt is the alt-text for the image, and <...>. You will need to replace these with the actual values.
  \end{minipage}}\\
    \midrule    
    \textbf{User 2}: & \textcolor{customorange}{\textbf{``Can you optimise the images on this article page for search engine optimization. Refactor the code accordingly by labeling the images or adding relevant image descriptions.''}} \\
    \textbf{Copilot:} & \colorbox{codebgd}{\texttt{imgAlt={imageDescription} // Add this line}}\\
    & \colorbox{codehlt}{\begin{minipage}{\linewidth}
    Sure, to optimize images for SEO, you can add alt attributes to your img tags. The alt attribute provides alternative information for an image if a user for some reason cannot view it (because of slow connection, an error in the src attribute, or if the user uses a screen reader).
  \end{minipage}}
    \\
  \bottomrule
\end{tabular}
    \label{tab:responses}
\end{table*}

\ipstart{AI Usage and Prompting Strategies}
Participants mainly used the autocomplete feature only when they had a clear mental model of the desired code structure and sought to accelerate the code typing process. In contrast, they heavily relied on the conversational interface for syntax inquiries, conceptual understanding, and the generation of code templates. We noticed that our participants wrote brief, task-oriented prompts that focused on immediate code solutions or specific interface modifications, often disregarding broader architectural considerations. Their prompting style was iterative and reactive, frequently requesting small incremental changes, fixes to previous outputs, or refinements to their vague prompts. 

Furthermore, none of the participants, including the two who were familiar with web accessibility, prompted with accessibility in mind. Instead, our participants' prompts centered around visual and functional attributes (e.g., `\textit{`add a gray background to the subscription form}'' (P4) or ``\textit{add a grey patch}'' (P1)). Consequently, the AI assistant's suggestions often failed to incorporate accessibility best practices automatically. Occasionally, our participants prompted for enhancements that indirectly aligned with accessibility requirements, and Copilot provided relevant accessibility suggestions, as shown in Table~\ref{tab:responses}. However, participants' overreliance on AI assistance often led them to assume that Copilot's code output was correct and complete. For instance, despite additional explanations from Copilot advising manual adjustments to image descriptions, participants directly pasted the code, resulting in code submissions with empty \colorbox{codebgd}{\texttt{alt}} attributes.

\begin{figure}
\includegraphics[width=0.49\textwidth,trim=80 65 50 30,clip]{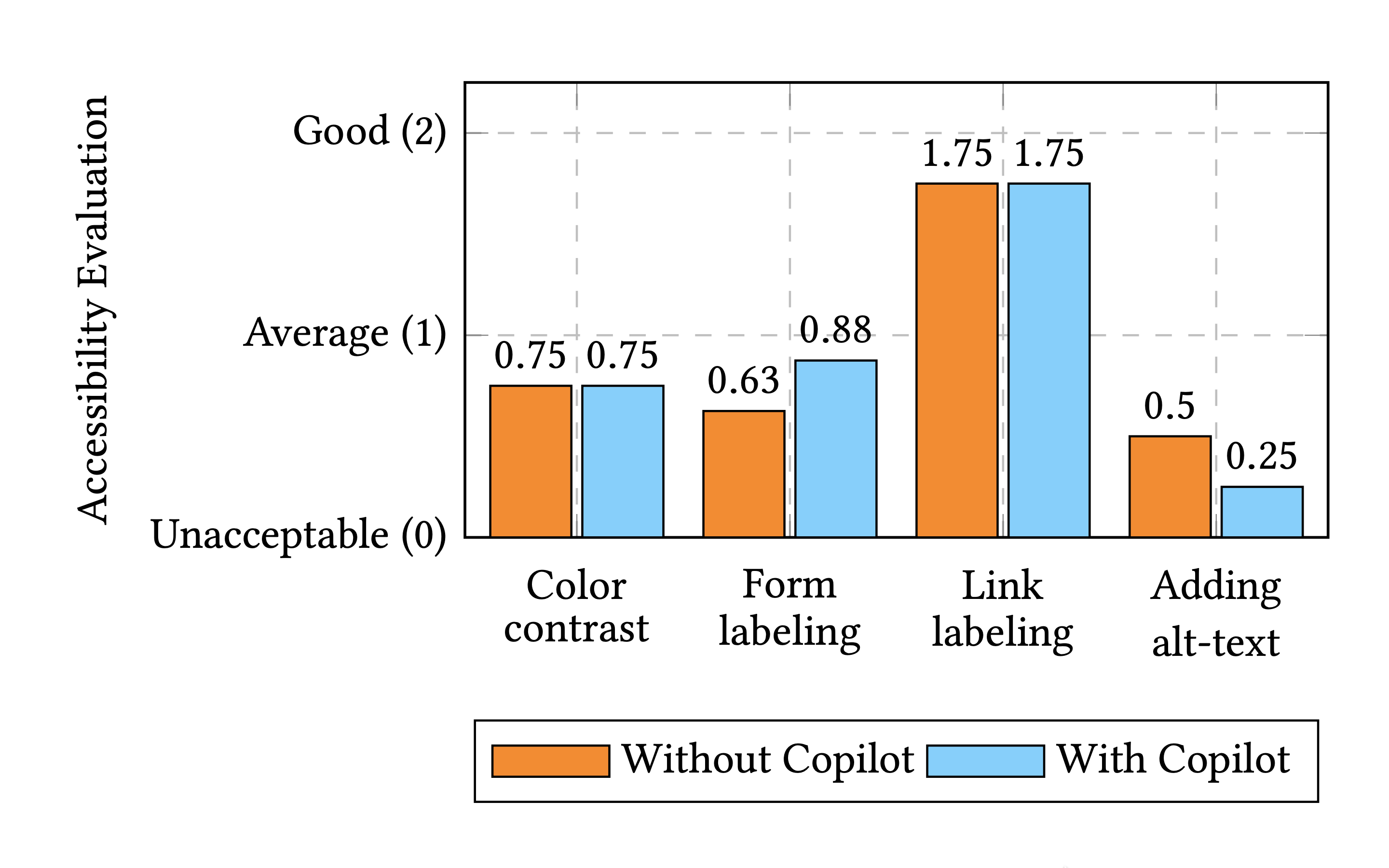}
\caption{Mean Accessibility Evaluation Scores by Tasks and Copilot Usage: Higher scores indicate success.}
\Description{The image shows a bar graph with mean scores for web accessibility tasks, comparing outcomes with and without GitHub Copilot. Scores range from 0 (Unacceptable), 1 (Average), 2 (Good). Adding alt-text on average worsened from 0.5 to 0.25 with the usage of Copilot. Button colour contrast and link labelling remained at 0.63 and 1.75 respectively regardless of Copilot use. Form labelling improved from 0.63 to 0.88 with Copilot usage.}
\label{manual-eval}
\end{figure}

\ipstart{Implications for Web Accessibility}
Our study showed mixed results of AI assistants in considering accessibility issues with no statistically significant difference between the experimental conditions, as shown in Figure~\ref{manual-eval}. Notably, Copilot could (sporadically) generate accessible components by utilizing patterns from other parts of a website. For example, it might automatically include proper labels for form fields, such as \colorbox{codebgd}{\texttt{<label for="email"> Email: </label>}} in a signup form. However, there were also instances where Copilot inadvertently introduced new accessibility issues. For example, when adding new button components with hover effects, it failed to ensure adequate contrast between the hover color and background. 

Further, the effectiveness of AI assistants was limited by the need for more sophisticated accessibility knowledge. Since our participants had limited awareness about these accessibility features, they would often ignore such suggestions by blindly accepting \colorbox{codebgd}{\texttt{alt = "" // Add your text here}} or manually deleting the \colorbox{codebgd}{\texttt{<label>}} tag. Some errors, such as providing blank alt-texts for informative images, were not even flagged by automated accessibility checkers because they interpret the image as decorative and consider this deliberate. This is particularly problematic as it implies that AI assistance might increase the risk of accessibility oversights, allowing critical errors to go unnoticed and uncorrected.

\section{Design Requirements}
\label{design-goals}
Our formative study identified three limitations in novice developers' interactions with AI assistants: (1) failing to prompt for accessibility considerations explicitly, (2) uncritically accepting incomplete code suggestions from Copilot, and (3) struggling to detect potential accessibility issues in their code. These shortcomings indicate possible directions to support accessibility through three design goals (\textbf{G1-G3}):

\ipstart{G1: Integrate System Prompts for Accessibility Awareness} 
Without explicit prompting, the AI assistant rarely produced accessibility-compliant code, reflecting the accessibility issues prevalent in its training data. However, it occasionally suggested accessibility features when participants indirectly prompted them, demonstrating its ability to recall accessibility practices from training data upon instruction. AI assistants should have a system prompt tuned towards following accessibility guidelines by default, for consistent generation of accessibility-compliant code, even when developers do not mention accessibility specifically. Further, the system prompt should also direct the assistant to suggest accessibility-focused iterative refinements.

\begin{highlight}
\ipstart{G2: Support Identification of Accessibility Issues}
\end{highlight}
Due to their unfamiliarity with accessibility standards, our participants were unable to identify compliance issues in the existing and modified code. They primarily prompted changes to individual components (such as buttons and forms), hardly addressing broader page-level accessibility concerns (such as heading structure or landmark regions). AI assistants should not only automatically generate accessibility-compliant code, but also provide real-time feedback to detect and resolve accessibility violations within the codebase. In addition, AI assistants and automated accessibility checkers should work in tandem to ensure that incomplete or incorrect implementations of the AI-suggested code are always flagged by the latter.

\begin{highlight}
\ipstart{G3: Encourage Developers to Complete AI-Generated Code}
\end{highlight}
Our observations revealed that accessibility implementation in AI-assisted coding workflows commonly required critical manual intervention to complete and validate AI-generated code. This involved replacing placeholder attributes, such as labels and alt-texts, with meaningful values and verifying color contrast ratios. However, we found that participants blindly copy-pasted code and proceeded further if there were no apparent errors. This behavior of deferring thought to suggestions has also been documented in previous work~\cite{mozannar2024reading}. To mitigate this, AI assistants should proactively remind developers to ensure that all necessary accessibility features -- such as contrast ratios or keyboard navigation support -- are fully implemented and verified.
\section{CodeA11y}
\label{system}
Guided by the design goals identified through our user study, we built CodeA11y, a GitHub Copilot Extension for Visual Studio IDE. In this section, we present the interactions that it supports and its system architecture.

    
CodeA11y has three primary features (\textbf{F1-F3}, aligned to G1-G3, respectively): (\textbf{F1}) it produces user interface code that better complies with accessibility standards, \begin{highlight}
(\textbf{F2}) it prompts the developer to resolve existing accessibility errors in their website, and (\textbf{F3}) it reminds the developer to complete any AI-generated code that requires manual intervention.
\end{highlight} CodeA11y is integrated into Visual Studio Code as a GitHub Copilot Extension\footnote{\url{https://docs.github.com/en/copilot/using-github-copilot/using-extensions-to-integrate-external-tools-with-copilot-chat}}, enabling CodeA11y to act as a chat participant within the GitHub Copilot Chat window panes. While we implemented this as an extension, it could be integrated directly into Copilot in the future.


\ipstart{Multi-Agent Architecture}
\begin{table*}
  \caption{Prompt instructions for the three LLM agents in CodeA11y}
  \label{tab:sys_prompt}
  \small
  \begin{tabular}{m{0.14\textwidth}|m{0.81\textwidth}}
    \toprule
    \textbf{Agent} & \textbf{Prompt Instruction Highlights}\\
    \midrule
    Responder Agent & {\begin{itemize}[leftmargin=*, partopsep=7pt]
        \item I am unfamiliar with accessibility and need to write code that conforms with WCAG 2.1 level AA criteria.
        \item Be an accessibility coach that makes me account for all accessibility requirements.
        \item Use reputable sources such as w3.org, webaim.org and provide links and references for additional learning. 
        \item Don't give placeholder variables but tell me where to give meaningful values.
        \item Prioritise my current request and don't mention accessibility if I give a generic request like "Hi".
    \end{itemize}} \\ \midrule
    Correction Agent & {\begin{itemize}[leftmargin=*, partopsep=7pt]
        \item Review the accessibility checker log and provide feedback to fix errors relevant to current chat context.
        \item If a log error relevant to current chat context occurs, provide a code snippet to fix it.
    \end{itemize}} \\ \midrule
    Reminder Agent & {\begin{itemize}[leftmargin=*, partopsep=7pt]
        \item Is there an additional step required by the developer to meet accessibility standards after pasting code?
        \item Reminder should be single line. Be conservative in your response, if not needed, say "No reminders needed."
        \item For example, remind the developer to replace the placeholder attributes with meaningful values or labels, or visually inspect element for colour contrast when needed.
    \end{itemize}}\\ 
    \bottomrule
  \end{tabular}
\end{table*}
CodeA11y has three LLM agents (Figure~\ref{fig:arch}): Responder Agent, Correction Agent, and Reminder Agent. We provide their prompt instruction highlights in Table~\ref{tab:sys_prompt}. These agents facilitate each of the above features (F1-F3) as follows:
\begin{itemize}
    \item \textbf{\textit{Responder Agent}} (for F1): This agent generates relevant code suggestions based on the developer's prompt. It assumes that the developer is unfamiliar with accessibility standards and automatically generates accessible code.
    The prompt instruction for this agent is adapted from GitHub's recommended user prompt for accessibility.\footnote{\url{https://github.blog/developer-skills/github/prompting-github-copilot-chat-to-become-your-personal-ai-assistant-for-accessibility/}}
    \item \textbf{\textit{Correction Agent}} (for F2): This agent parses through accessibility error logs produced by an automated accessibility checker (axe DevTools Accessibility Linter\footnote{\url{https://www.deque.com/axe/devtools/linter/}}) to
    hint the developer at making additional accessibility fixes in the component or page being currently discussed in the chat context.
    \item \textbf{\textit{Reminder Agent}} (for F3): This agent reviews the Responder Agent's suggestions and identifies required manual steps for completing their implementation. It accordingly sends reminder notifications to the developer through the Visual Studio IDE infrastructure.
\end{itemize}
\begin{highlight}
The agents are provided with several different sources of context:
\begin{itemize}
\item \textbf{Code Context}: the 100 lines of code centered around the cursor position in the active files.
\item \textbf{Chat Context}: the current active chat window interaction.
\item \textbf{Accessibility Linter Logs}: automated Axe DevTools Accessibility Linter error logs, refreshed periodically.
\item \textbf{Project Context}: code context from the \colorbox{codebgd}{\texttt{README}} and \colorbox{codebgd}{\texttt{index}} files, which contain information about the web framework that is being used, information about project structure, and other key configuration details.
\end{itemize}

Due to the constraints in the context window, we optimized our prompts and filtered this context when it exceeded 4000 characters. The agents use GPT-4o as the back-end model from the same OpenAI GPT family of models that powers GitHub Copilot.
\end{highlight}

\begin{highlight}
\begin{figure}
 \includegraphics[width=0.45\textwidth, trim=0 0 405 0, clip]{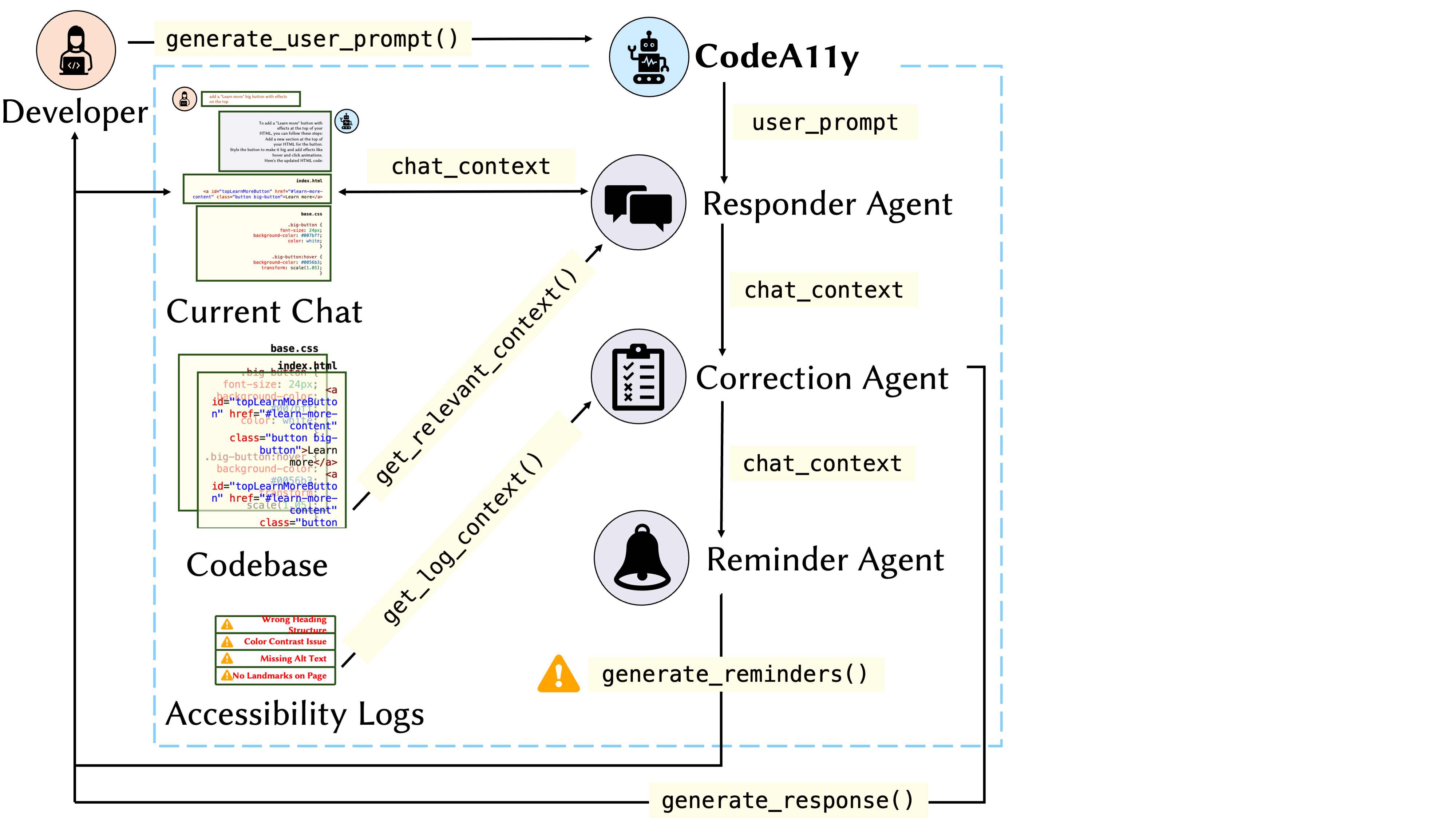}
    \caption{CodeA11y Architecture: Multi-agent workflow}
    \Description{CodeA11y contains three agents: Responder Agent, Correction Agent and Reminder Agent which take the user_prompt, code_context and log_context respectively to generate a response for the chat window.}
    \label{fig:arch}
\end{figure}
\end{highlight}

\begin{figure*}
    \includegraphics[width=\textwidth]{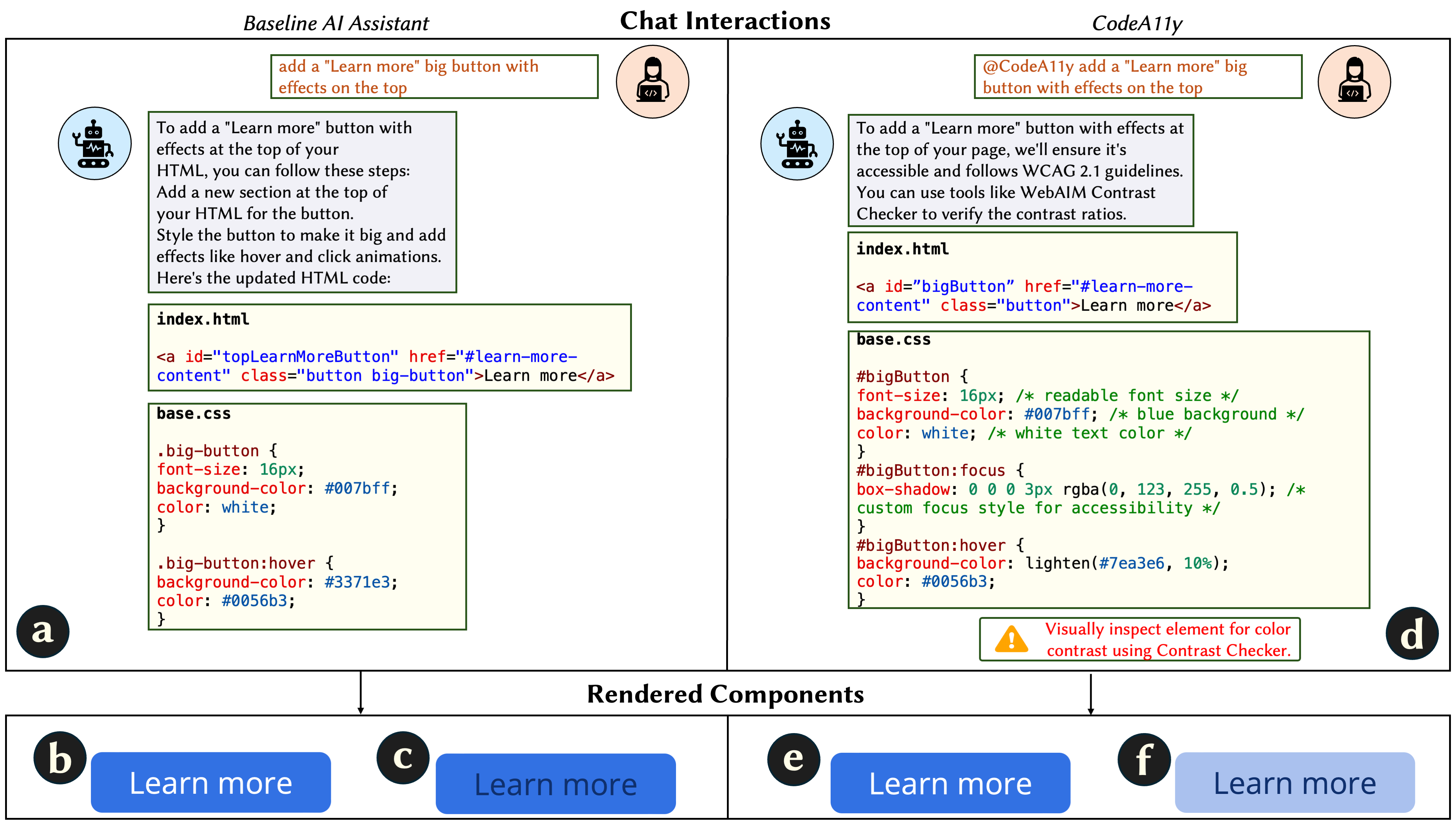}
    \caption{Contrasting responses for the same task across AI-assistants, \begin{highlight}showing differences in workflows. Developers had access to both the code and the rendered user interface.\end{highlight} (a) and (d) represent conversations with the baseline assistant, and CodeA11y respectively. (b) and (e) show the buttons generated by each assistant in their default state. (c) and (f) display the buttons when hovered over, illustrating the differences in button color contrast.}
    \Description{The notable differences in the interactions are as follows: CodeA11y reinforces accessibility practices in the text response, generates a reminder to check for button contrast and adds appropriate colours in the code. This results in a button with appropriate color contrast in both default and hover states. Conversely, the baseline assistant gives code for a button with the same colors in the default state, but significantly poorer color contrast in the hover state.}
    \label{fig:interaction}
\end{figure*}

\ipstart{User Interaction}
Developers invoke\footnote{In the long term, the goal is for GitHub Copilot to invoke CodeA11y automatically during frontend development tasks.} CodeA11y in the GitHub Copilot Chat window panes (includes Quick Chat and Chat View) using \colorbox{codebgd}{\texttt{@CodeA11y}}. When a developer prompts CodeA11y, an internal \colorbox{codebgd}{\texttt{chat\_context}} state is established, storing the latest user prompts and agent responses. The \colorbox{codebgd}{\texttt{get\_relevant\_context()}} function is called, \begin{highlight}
which passes the source code and project context\end{highlight} to the Responder Agent. The agent generates code suggestions, accessibility explanations, and links to additional resources and updates \colorbox{codebgd}{\texttt{chat\_context}}. The \colorbox{codebgd}{\texttt{get\_log\_context()}} function is called, \begin{highlight}
which passes the accessibility linter logs\end{highlight} to the Correction Agent. This agent 
suggests additional fixes and displays the responses in the chat pane. Lastly, the updated \colorbox{codebgd}{\texttt{chat\_context}} state is forwarded to the Reminder Agent, which generates and sends reminder notifications. Figure~\ref{fig:interaction} illustrates a typical interaction between a developer and CodeA11y, showing how it compares to baseline assistants like GitHub Copilot.

\section{User Evaluation}
\label{eval}

\begin{table*}
  \caption{The distribution of opinions on AI-powered programming tools and their awareness of web accessibility based on the responses from participants in the evaluation study. The percentages in the distribution column indicate the proportion of participants who either disagree (including `strongly disagree', `disagree' and `slightly disagree') or agree (including `strongly agree', `agree' and `slightly agree') with the provided statements.}
  \label{tab:aware2}
  \begin{tabular}{>{\raggedright\arraybackslash}p{0.6\textwidth}|>{\raggedright\arraybackslash}p{0.35\textwidth}}
    \toprule
    \textbf{Statement} & \textbf{Distribution}\\
    \midrule
    ``I trust the accuracy of AI programming tools.''& \likertpctt{0}{1}{2}{3}{10}{4}{0}\\
    ``I am proficient in web accessibility.''& \likertpctt{5}{5}{4}{1}{4}{0}{1}\\
    ``I am familiar with the web accessibility standards, such as WCAG 2.0.''& \likertpctt{11}{4}{1}{1}{0}{1}{2}\\
    ``I am familiar with ARIA roles, states, and properties.''& \likertpctt{13}{4}{0}{1}{0}{1}{1}\\
  \midrule
\end{tabular}
\begin{tabular}{@{}>{\centering\arraybackslash}p{\textwidth}@{}}
        \textcolor{customorange}{\rule{7pt}{7pt}} Strongly Disagree \;
        \textcolor{custommidorangepeach}{\rule{7pt}{7pt}} Disagree \;
        \textcolor{custompeach}{\rule{7pt}{7pt}} Slightly Disagree \;
        \textcolor{customgray}{\rule{7pt}{7pt}} Neutral \;
        \textcolor{customlightblue}{\rule{7pt}{7pt}} Slightly Agree \;
        \textcolor{custommidbluelightblue}{\rule{7pt}{7pt}} Agree \;
        \textcolor{customblue}{\rule{7pt}{7pt}} Strongly Agree \\ 
    \bottomrule
    \end{tabular}
\end{table*}

We conducted a within-subjects user study with 20 new participants to evaluate CodeA11y's effectiveness in guiding novice developers toward adhering to accessibility standards, as compared to Copilot.

\subsection{Methodology}
We made the following revisions to our formative study protocol (Section~\ref{form_methods}). First, the experimental conditions were updated as follows: (1) the control condition involved using the baseline AI assistant (GitHub Copilot), and (2) the test condition where the participants used CodeA11y. Second, we changed the post-task survey to a brief semi-structured interview to get more nuanced insights about the usability of our system. \begin{highlight} We analyzed interview responses to better understand the factors shaping participants' assistant preferences and their perceptions of any new coding practices introduced during the study.\end{highlight} Third, we used Visual Studio Code as the IDE interface (which had advanced AI updates since the formative study -- regarding both model performance and introduction of new features such as Inline Chat). Finally, we recruited 20 new participants for this subsequent study, with no prior exposure to the formative study, to evaluate CodeA11y's \begin{highlight}
performance on the same UI development tasks\end{highlight}. These participants were the same demographic as our formative participants (students; multiyear programming experience; 6 female and 14 male; ages ranged from 22 to 30). Again, most participants were unfamiliar with the web accessibility standards (Table~\ref{tab:aware2}), but most (90\%) had experience using AI programming tools. The IRB approved all our modifications.

\begin{table*}
  \caption{The distribution of participants' opinions on GitHub Copilot and CodeA11y, as well as their ease of completing tasks with these tools. The distribution column shows the count of responses from Strongly Disagree (1) to Strongly Agree (7).}
  \label{tab:satisfaction}
  \begin{tabular}{>{\raggedright\arraybackslash}p{0.5\textwidth}|>{\raggedright\arraybackslash}p{0.45\textwidth}}
    \toprule
    \textbf{Statement} & \textbf{Distribution}\\
    \midrule
    \multicolumn{1}{l}{``I am satisfied with the code suggestions provided by'':} & \\ \midrule
    GitHub Copilot & \likerteval{1}{1}{2}{1}{8}{3}{4} \\
    CodeA11y & \likerteval{1}{2}{0}{2}{4}{7}{4} \\ \midrule
    \multicolumn{1}{l}{``I found it easy to complete the coding tasks with'':} & \\ \midrule
    GitHub Copilot & \likerteval{0}{3}{0}{1}{3}{8}{5} \\
    CodeA11y & \likerteval{0}{2}{0}{0}{3}{8}{7} \\
    \midrule
\end{tabular}
\begin{tabular}{@{}>{\centering\arraybackslash}p{\textwidth}@{}}
    \textcolor{customorange}{\rule{7pt}{7pt}} Strongly Disagree \;
    \textcolor{custommidorangepeach}{\rule{7pt}{7pt}} Disagree \;
    \textcolor{custompeach}{\rule{7pt}{7pt}} Slightly Disagree \;
    \textcolor{customgray}{\rule{7pt}{7pt}} Neutral \;
    \textcolor{customlightblue}{\rule{7pt}{7pt}} Slightly Agree \;
    \textcolor{custommidbluelightblue}{\rule{7pt}{7pt}} Agree \;
    \textcolor{customblue}{\rule{7pt}{7pt}} Strongly Agree \\ 
    \bottomrule
\end{tabular}
\end{table*}
To avoid biasing participants towards adhering to accessibility guidelines, we did not disclose the specific purpose of the CodeA11y plugin. For the duration of the study, we renamed the assistant ``Codally'' and described it as a general-purpose chat assistant for website editing. We assumed the interface would be intuitive, similar to widely used assistants, and therefore briefed participants only on basic AI assistant usage (e.g., Copilot), deliberately withholding explanations of error pop-ups to prevent influencing their behavior before the main study tasks. However, during the course of our study, we realized that VS Code was dismissing popup boxes created by our plugin more rapidly than expected - causing some participants to miss them. \begin{highlight} After 8 participants, we switched from floating popups to modals (which prevent the IDE's auto-dismissal) due to a technical limitation. Both notification strategies do not require users to address errors, making them valid design choices. In our baseline comparison, we aggregate data from all users and include anecdotal observations of user behavior with each strategy. We acknowledge that such UI design choices may introduce variability and plan to investigate this further in future work.
\end{highlight} 
\definecolor{highlightyellow}{RGB}{255, 253, 232}
\definecolor{highlightbleu}{RGB}{18, 96, 157}
\definecolor{shade2}{RGB}{0,0,120}
\definecolor{customdarkred}{RGB}{150, 0, 0} 
\definecolor{customred}{RGB}{210, 70, 70}   
\definecolor{custommidred}{RGB}{225, 150, 150} 
\definecolor{customneutral}{RGB}{200, 200, 200} 
\definecolor{custommidblue}{RGB}{90, 180, 225}  
\definecolor{customblue}{RGB}{51, 130, 185}   
\definecolor{customdarkblue}{RGB}{12, 80, 145} 
\subsection{Results}
Here, we present the results of our subsequent evaluation study.

\begin{figure}
\centering
\includegraphics[width=0.49\textwidth,trim=40 60 50 50,clip]{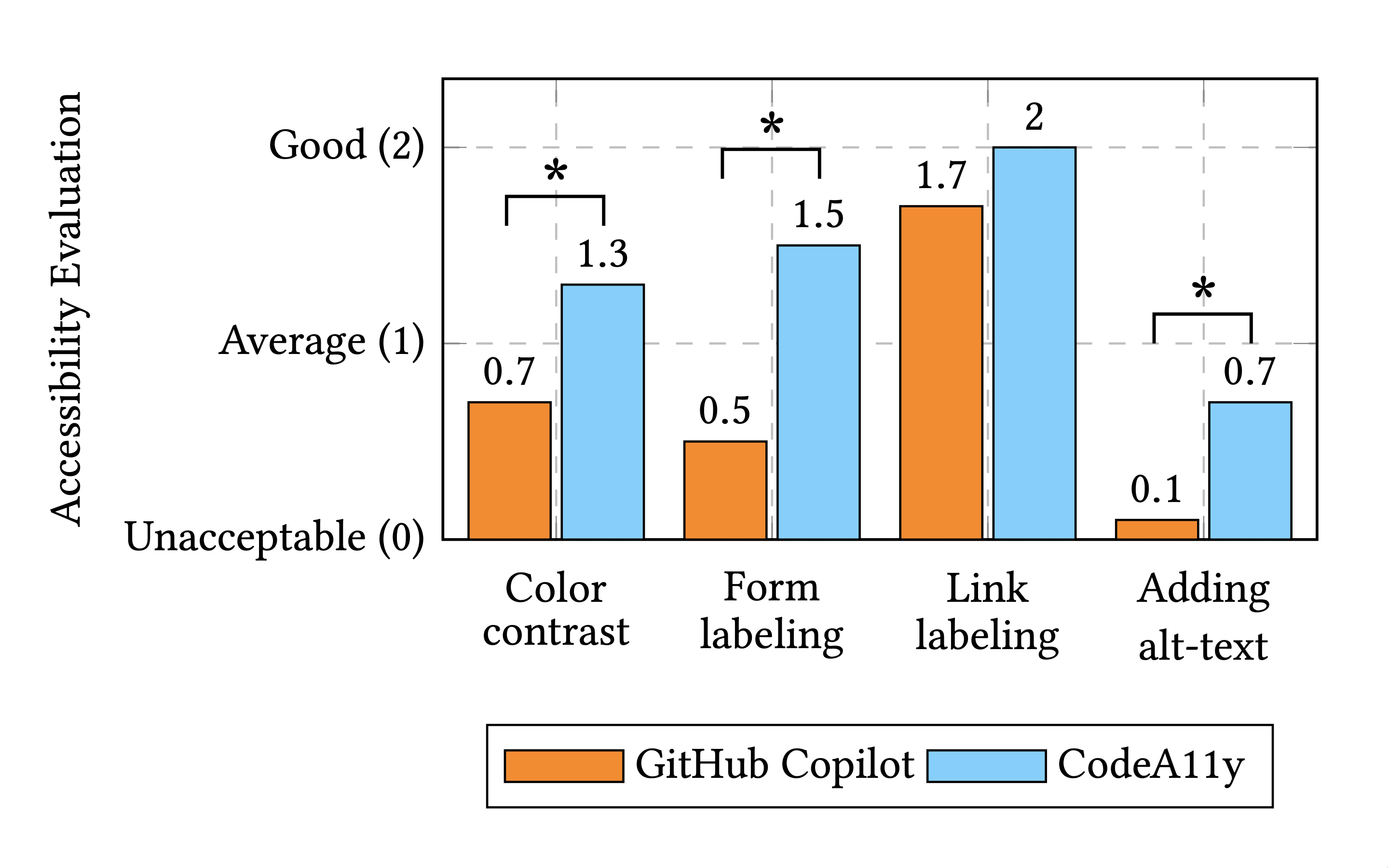}
\caption{Mean Accessibility Evaluation Scores by Tasks and AI Assistant: Higher scores indicate success.}
\Description{The image shows a bar graph with mean scores for web accessibility tasks, comparing outcomes with GitHub Copilot and CodeA11y AI assistants. Scores range from 0 (Unacceptable), 1 (Average), 2 (Good). With the usage of CodeA11y, button color contrast, form labelling and adding alt texts improved with statistical significance, from 0.7 to 1.3; 0.5 to 1.5; and 0.1 to 0.7, respectively. Link labeling also improved, though not with statistical significance, from 1.7 to 2.}
\label{sec-eval}
\end{figure}

\ipstart{Accessibility Improvements}
\begin{highlight}We implemented the accessibility assessments using the same measures outlined in our formative study (Table~\ref{tab:tasks}).\end{highlight} Notably, our participants demonstrated a marked improvement in generating accessible web components and resolving accessibility issues with CodeA11y (Figure~\ref{sec-eval}). CodeA11y facilitated the automatic addition of form labels and ensured contrasting colors for button states, leading to statistically significant enhancements in accessibility outcomes. Specifically, participants performed better at adding form labels ($\mu=1.5$, $\sigma=0.85$) compared to GitHub Copilot ($\mu=0.5$, $\sigma=0.85$; $t=2.63$, $p<0.05$) and in ensuring contrasting button colors ($\mu=1.3$, $\sigma=0.67$ vs. $\mu=0.7$, $\sigma=0.82$; $t=1.78$, $p<0.05$). We also observed improvements in adding alt-texts with CodeA11y ($\mu=0.7$, $\sigma=0.95$ vs. $\mu=0.1$, $\sigma=0.32$; $t=1.9$, $p<0.05$). Though we did not find any statistical improvements in labeling links (perhaps because GitHub Copilot did a decent job at this task itself), all participants who used CodeA11y successfully completed this task ($\mu=2$, $\sigma=0$ vs. $\mu=1.7$, $\sigma=0.67$; $t=1.9$, $p=0.09$).

\ipstart{Developers' Perspectives}
Overall, participants reported no statistically significant difference in satisfaction ($\mu=5.15$, $\sigma=1.75$ vs. $\mu=4.95$, $\sigma=1.67$; $t=0.37$, $p=0.36$) and ease of use ($\mu=5.8$, $\sigma=1.47$ vs. $\mu=5.4$, $\sigma=1.67$; $t=0.8$, $p=0.21$) between CodeA11y and Github Copilot, respectively, as illustrated in Table~\ref{tab:satisfaction}.

During the post-study interviews, participants provided additional reasoning for their preferences. Most (n=16) participants \begin{highlight} did not have a specific preference between the two assistants, which is consistent with the conclusion of our statistical analysis. 
Others did indicate a preference (n=3) but provided reasoning that was based on the complexity of the task rather than assistant features, ``\textit{I liked the first assistant (CodeA11y) better, maybe because of the tasks. The second one (GitHub Copilot) required me to understand the code, and the first directly gave me the code. That’s the difference.}'' (\textbf{P18}) \end{highlight}

\begin{highlight} We asked our participants if they were introduced to any new coding practices by either of the assistants. To our surprise, only 4 participants mentioned accessibility, demonstrating CodeA11y's effectiveness in ``silently'' improving the accessibility of our participants' UI code. \end{highlight} These participants noted that they had not these considerations before. However, some mentioned either not paying attention to them or subconsciously rejecting them, as they were primarily focused on completing the tasks, which they perceived to be unrelated to accessibility:
\begin{quote}
    ``\textit{I did not find any difference (between the assistants). When I was prompting CodeA11y, it was hinting at me to use alt texts, which was not happening in Copilot. It didn’t come to me by default, so that was good ... But I don't think I implemented that.}'' (\textbf{P27})
\end{quote}

Still, they appreciated CodeA11y for emphasizing best practices for accessibility. For instance, \textbf{P27} continued: ``\textit{I did see a few of the popups, and they did mention some interesting points like you need to consider the color of the button when you add a new button because if people are color blind, they might not be able to notice it.}'' \begin{highlight} Further, \textbf{P19}, familiar with web accessibility but not proficient, realized that although he did not learn anything new about CodeA11y's color contrast suggestion, he noticed a visible difference in user experience after accepting it. Our sole participant who claimed proficiency in accessibility valued support for a specific framework:\end{highlight}
\begin{quote}
    ``\textit{I am familiar with accessibility coding practices but not in the React Native environment; I don't know if I would have needed that help in HTML, but I liked that it tried to highlight accessibility practices in React Native.}'' (\textbf{P21})
\end{quote}

\ipstart{Other Observations} 
Participants frequently made minor edits to the AI-generated code for refining the visual appearance of web components. For instance, \textbf{P18} remarked, ``\textit{CodeA11y did the job for me; I only had to change the property values.}'' However, while focusing on visual adjustments, participants occasionally removed accessibility enhancements suggested by CodeA11y. Further, many participants still overlooked the manual validation steps required for implementing more advanced accessibility features.
Participants appeared to consistently lack interest in the floating popups, so 50\% of participants using CodeA11y still added uninformative alt-texts. We observed a slightly different pattern with the modal reminders. Our participants initially paid attention to the modal interface, but then began to just close them. Although performance across the two reminder types is hard to assess objectively due to the small sample size, we observed higher means for the alt-text and form label tasks for the modal reminders. As a whole, the reminders proved somewhat effective: none of the participants using CodeA11y submitted empty alt-texts, which meant at least automated accessibility checkers would not consider the images decorative.

\section{Discussion}
\label{discuss}

This paper has explored how AI coding assistants currently contribute to UI code that is accessible to people with disabilities. While these tools offer a new opportunity for achieving accessibility, we have revealed the remaining challenges and showed how they could be addressed with changes to the way the coding assistants operate.

\ipstart{Which comes first: Adoption or Awareness?}
Adoption is a perennial challenge faced by most accessibility technologies, even when the technology could lead to substantial improvements in user experience.
One reason adoption is low is because awareness is low -- people who could benefit from access technology do not know about it. For example, a survey found that only 10\% of older adults knew what the term ``accessibility'' meant and therefore did not enable any useful settings~\cite{wu2021can,peng2019personaltouch}.
Similarly, developers benefit in many ways from tools that improve the accessibility of their code (\textit{e.g.,} linters, scanners), resulting in better-designed applications and reaching more users. Unfortunately, many developers are unaware of these tools or unwilling to adopt new practices that require changing their original workflow.
For example, while AI assistants like Copilot are capable of generating accessible code, our formative study found that developers were unaware or unwilling to explicitly prompt it to do so.

One goal of our work was to investigate whether developers could increase the adoption of accessibility technology and development practices independently or in tandem with awareness.
For example, prior work~\cite{bigham2014making} found that by ``opportunistically'' zooming into web pages and configuring settings, users could automatically benefit from improved accessibility.
Our motivation is similar: we aimed to improve code accessibility while introducing minimal changes to existing AI-assistant developer workflows \textit{i.e.,} GitHub Copilot.
According to the Visual Studio Marketplace, GitHub Copilot has been installed over 20 million times (as of the time of writing), suggesting that many developers are already familiar with the plugin's interactions, tooling, and interface. 
Our results show that CodeA11y significantly improved code accessibility while maintaining a similar (slightly improved) ease of use to Copilot.
This suggests that if GitHub Copilot included our set of features or were willing to use a similar plugin without requiring substantial deviation from existing workflows, millions of developers could start writing more accessible code immediately.




\ipstart{AI-Assisted but Developer-Completed}
Even when developers adopt accessibility tools, additional expertise is required to maximize their utility.
This is especially true for AI assistants, which are incapable of generating entirely correct or accessible code. Our work offers some insight into the manual effort needed to write accessible code.
Both our formative and evaluation studies underscore the necessity for developers to manually intervene with AI-generated code to effectively implement accessibility features. A recurrent challenge in AI-assisted coding is the generation of incomplete or boilerplate code, which often requires developers to take additional steps for completion and validation. Our findings reveal that novice developers tend to critically evaluate AI outputs in areas they prioritize, such as visual enhancements, while overlooking aspects they are less familiar with, like accessibility.

One of the tensions of this work is that while we aim to increase the adoption of AI-driven accessibility tools among users with little expertise or awareness, some degree of understanding is required to use these tools effectively. CodeA11y and other tools can employ strategies to scaffold this interaction, \textit{e.g.,} asking a user ``can you describe what's in this image?'' instead of asking them directly for ``alternative text.'' However, developers ultimately need to be willing to expend additional effort and manually implement the more challenging aspects of this work. Thus, while our work suggests that it is possible to ``silently'' improve the accessibility of developer-written code, it is ultimately not a replacement for better accessibility awareness, development practices, and education. Nevertheless, CodeA11y could help gradually improve awareness by slowly introducing and explaining accessibility concepts to users after they have found benefits from using the tool.

\ipstart{Limitations \& Future Work}
We describe several limitations in the current scope of the study and identify avenues for future work to build upon our findings:

First, the utility of the CodeA11y plugin was limited by the constraints placed by our target development environment (Visual Studio Code). 
Because CodeA11y was implemented as a Copilot plugin, we could only access a few APIs available to standard VSCode IDE plugins. The Copilot plugin infrastructure was limiting because it restricted the source code that could be passed to the model (\textit{i.e.,} context window length). Our implementation contained some mechanisms for heuristically determining the most relevant files but ultimately serves as a proof of concept of what would be possible in the future, well-integrated version (\textit{e.g.,} built into Copilot). These factors affected the code generation of our system.

\begin{highlight}
Second, although our user study provides statistical evidence that CodeA11y helps developers write more accessible website code, we acknowledge certain limitations of AI coding assistants that could affect their overall effectiveness and reliability. One well-documented issue, particularly with proactive AI assistants~\cite{chen2024need}, is their potential to provide untimely or irrelevant guidance. For instance, the models may occasionally suggest fixes for problems that do not exist (i.e., false positives), such as recommending changes to code that is already fully compliant with accessibility standards.
While our study did not surface such occurrences---likely because CodeA11y’s suggestions were tied directly to verified issues from an accessibility checker, rather than the tool identifying issues on its own---this risk becomes more salient as AI assistants evolve to more proactively identify and address accessibility problems. Still, prior research suggests that developers often tolerate false positives more readily than false negatives~\cite{kocielnik2019will}, reasoning that overly cautious guidance from an assistant is less harmful than failing to flag genuine accessibility issues. Indeed, even standalone accessibility checkers, which are also known for producing false positives~\cite{huq2023a11ydev}, have been widely adopted due to their overall beneficial effect on UI quality.
Moreover, the occasional presence of false positives does not necessarily negate the value of employing such tools. By raising awareness and prompting developers to consider accessibility from the outset, AI assistants can help cultivate a proactive mindset toward inclusive design. In this sense, the technology does not need to achieve perfect accuracy to have a net positive effect. As the underlying models and APIs improve, and assistants become better integrated with real-world workflows, their precision and utility in improving accessibility are likely to increase.

Third, while our study demonstrates the potential of CodeA11y to encourage developers to adopt accessibility practices, we acknowledge that the broader impact of AI coding assistants on long-term learning and behavior changes (e.g., for accessibility awareness) remains underexplored in the current scope of the study. Research on how real-time AI tools can help developers internalize new practices, such as accessibility, or foster long-term behavioral changes could have strengthened the case for CodeA11y’s instructional components. For instance, prior work has shown that AI coding tools can enhance immediate task performance but may not consistently lead to deeper learning or sustained skill retention~\cite{kazemitabaar2023studying}. Similarly, studies on meta-cognitive demands in AI-assisted workflows emphasize the importance of tools promoting reflective learning and adaptive strategies, particularly as developers integrate them into their daily practices~\cite{tankelevitch2024metacognitive}. Although our findings suggest that CodeA11y has the potential to raise awareness of accessibility issues through direct integration with verified accessibility checks, further research is needed to understand whether such tools can foster a lasting developer's commitment to accessibility or similar best practices. Additionally, exploring how these tools impact broader developer workflows, collaboration habits, and the ability to generalize learned behaviors across contexts would provide a more comprehensive view of their instructional value. By examining these dimensions, future studies could better elucidate the role of AI coding assistants in shaping not just productivity, but also the culture of inclusive and responsible software development. While CodeA11y focuses on improving accessibility, its approach could extend to other non-functional requirements, such as privacy and security. Investigating how specialized copilots could be seamlessly invoked within mainstream coding assistants for high-stakes scenarios---such as leveraging CodeA11y for front-end development tasks---represents a promising direction.
\end{highlight}

Finally, we see opportunities to iterate on and refine CodeA11y’s design. Our conservative approach adhered closely to Copilot’s existing interface to minimize friction during adoption. Future work could explore how developers respond to new features and interactions, identifying areas where innovation could enhance usability and functionality without compromising adoption.
\begin{highlight}
By addressing current limitations and exploring broader applications, tools like CodeA11y can refine how developers approach accessibility and other critical non-functional requirements. Beyond technical improvements, such advancements hold the potential to redefine AI’s role in shaping more inclusive, secure, and efficient coding practices. 
\end{highlight}


\section{Conclusion}
\label{conc}
Our work bridges decades of accessibility efforts with AI coding assistants, offering a novel solution to persistent web accessibility challenges. 
Through a formative user study, we identify shortcomings in how current AI-assisted development workflows handle accessibility implementation. Accordingly, we develop CodeA11y, a GitHub Copilot Extension, and demonstrate that novice developers using it are significantly more likely to create accessible interfaces. By focusing on integrating accessibility improvements seamlessly into everyday development workflows, this work marks a first step toward fostering accessibility-conscious practices in human-AI collaborative UI development.

\bibliographystyle{ACM-Reference-Format}
\bibliography{paper}

\appendix









\end{document}